\documentclass[%
 reprint,
superscriptaddress,
 amsmath,amssymb,
prb,
longbibliography,
numerical,
]{revtex4-1}

\usepackage{graphicx}
\usepackage{dcolumn}
\usepackage{bm}
\usepackage[normalem]{ulem}
\usepackage{color}
\usepackage{hyperref}

\newcommand{\bra}[1]{\left\langle #1 \right|}
\newcommand{\ket}[1]{\left| #1 \right\rangle}

\newcommand{\Hamiltonian}{{\cal H}} 
\newcommand{\id}{{\normalfont\hbox{1\kern-0.15em \vrule width .8pt depth-.5pt}}}

\newcommand{\pd}[2]{\frac{\partial #1}{\partial #2}}

\newcommand{\hide}[1]{}

\newcommand{\half}{\frac{1}{2}}
\newcommand{\ad}{a^\dagger}

\newcommand{\alphas}{\alpha^*}
\newcommand{\dalpha}{\frac{\partial}{\partial\alpha}}
\newcommand{\dalphas}{\frac{\partial}{\partial\alphas}}

\begin{document}
\title{Quantum dynamics in a tiered non-Markovian environment}

\author{Amir Fruchtman}
\affiliation{Department  of  Materials,  University  of  Oxford,  Oxford  OX1  3PH,  United  Kingdom}
\author{Brendon W. Lovett}
\affiliation{SUPA, School of Physics and Astronomy, University of St Andrews, KY16 9SS, United Kingdom}
\author{Simon C. Benjamin}
\affiliation{Department  of  Materials,  University  of  Oxford,  Oxford  OX1  3PH,  United  Kingdom}
\author{Erik M. Gauger}
\affiliation{Centre  for  Quantum  Technologies,  National  University  of  Singapore,  3  Science  Drive  2,  Singapore  117543}
\affiliation{Department  of  Materials,  University  of  Oxford,  Oxford  OX1  3PH,  United  Kingdom}
\affiliation{SUPA, Institute of Photonics and Quantum Sciences, Heriot-Watt University, Edinburgh EH14 4AS, United Kingdom}

\begin{abstract}
We introduce a new analytical method for studying the open quantum systems problem of a discrete system weakly coupled to an environment of harmonic oscillators. Our approach is based on a phase space representation of the density matrix for a system coupled to a two-tiered environment. The dynamics of the system and its immediate environment are resolved in a non-Markovian way, and the environmental modes of the inner environment can themselves be damped by a wider `universe'. Applying our approach to the canonical cases of the Rabi and spin-boson models we obtain new analytical expressions for an effective thermalisation temperature and corrections to the environmental response functions as direct consequences of considering such a tiered environment. A comparison with exact numerical simulations confirms that our approximate expressions are remarkably accurate, while their analytic nature offers the prospect of deeper understanding of the physics which they describe. A unique advantage of our method is that it permits the simultaneous inclusion of a continuous bath as well as discrete environmental modes, leading to wide and versatile applicability.

\end{abstract}

\date{\today}

\pacs{Valid PACS appear here}
\maketitle

\section{Introduction}

The field of open quantum systems, originally devised for quantum optics problems, has recently gained significant traction in the study of condensed matter systems: This is due to the exquisite level of quantum control that is becoming available over increasingly mesoscopic solid state systems, as well as the tantalising prospect that Nature itself may be harnessing quantum effects under adverse `warm and wet' conditions, e.g. in photosynthesis\cite{Engel2007,Collini2010} and the avian compass\cite{Cai2010, Gauger2011}.
In current literature there is a range of methods to evaluate the evolution of a general open quantum system, from the straightforward but approximate weak-coupling master equation approach~\citep{breuer2007theory} through to the fully-numerical path integral based on quasi-adiabatic propagator path integral (QUAPI)~\cite{Makri1992,Makri1995,Makri1995a,Nalbach2010,Nalbach2011b}. It is important to find ways of treating quantum systems embedded in environments that are realistically complex, both in terms of their structure and their non-Markovian nature (i.e. environments which have a `memory'). When a new approach is analytic rather than numerical, there is the considerable benefit that one gains a route to intuitive insight as well as a simulation tool.

In this paper we introduce a method based on a sequence of three steps: First, we introduce the `{\it P} matrix', which allows a phase space description of a multilevel system coupled to complex environment. Second, we perform a perturbative expansion of the resulting dynamical solution. Finally, we express the reduced dynamics in terms of an influence functional, a quantity which allows new insights into the behaviour of open systems.
Our method is intuitive, highly accurate as long as the system environment coupling does not get too large, and works for general spectral densities. 

In contrast to many conventional open quantum system approaches, such as those mentioned above, we consider a hierarchical environment consisting of two tiers. The outer tier represents a zero-correlation-time heat bath that acts on an inner tier that is the immediate environment of the system. The inner tier may consist of a single harmonic oscillator, a continuous bath of oscillator modes, or any additive combination thereof. 

Previous works such as Refs.~\onlinecite{Breuer2004,Imamog1994,Gassmann2002} consider similarly tiered environments for a different conceptual reason: in those cases a single environmental tier is subdivided with the purpose of capturing more accurate, non-Markovian dynamics. In a similar manner, Ref.~\onlinecite{Katz2008} considers a second tier which is constantly randomized for gaining a numerical advantage in simulating a singly tiered environment. By contrast, our approach here is not motivated by `mathematical' convenience but is rather designed to capture a commonly occurring `physical' reality. This latter motivation had already been applied to some specific models such as the damped Jaynes-Cummings model \cite{Scala2007,Wolf2008} and fictitious harmonic oscillators~\cite{Imamog1994}, and the idea has led to the theory of pseudo-modes\cite{Dalton2001} (intrinsically restricted to zero temperature). 
A similar idea underlies the so-called `reaction coordinate' method, where the inner tier is a single harmonic oscillator that is coupled to a wider environment\cite{Garg1985,Thorwart2000,Hughes2009}, an approach that is often referred to as a `structured environment' in the literature\cite{Thorwart2004,Goorden2004,Goorden2005,Huang2008}. This method employs a mapping between the original environment and a spin boson model with an effective spectral density\cite{Garg1985}.

The method we introduce here applies to a general choice of system and bosonic environment at finite temperature, and the two environmental tiers typically represent different environmental influences. There also exist methods for modelling a long or infinite chain of identical environmental tiers,  for example, the problem of a quantum system coupled to the end of a linear chain of fermions~\cite{Rossini2007} or bosons~\cite{Woods2014}. We remark that  our method  remains applicable when there is no natural division into separate tiers and only a single environment is considered (or when both tiers arise from the same environment). In this case we still obtain non-Markovian contributions to the dynamics, and when applied to canonical cases, we recover known results from the literature. However, our method is more distinctive when two different environmental influences are present.

Another active area of research on open quantum systems is that of hierarchical equations, which was pioneered by Tanimura~\cite{Tanimura1989,Tanimura1990,Tanimura1991} in the late 80's. This   includes hierarchical equations for both the density matrix ~\cite{Ishizaki2009a,Ishizaki2009,Makri1995b,Ishizaki2005} and wavefunction \cite{Suß2014},  generally relying on a specific form of the memory kernel of the bath. Non-Markovian state quantum state diffusion~\cite{Roden2011,Mary1997,Diosi1998} also makes use of a hierarchy of abstract functionals and has recently been used to study energy transfer in molecular aggregates~\cite{Ritschel2011b}. Note, however, that the technique presented in this paper is conceptually quite different from any of these hierarchical approaches, since our interest focusses on a doubly tiered physical  environment instead of mathematical hierarchies of equations. 

Our approach of using a two-tiered environment makes our technique particularly suitable for modelling several of today's most intensely studied experimental systems: This includes many examples of discrete quantum systems interacting with an optical or mechanical resonator, such as, e.g., NV$^{-}$ centres on diamond cantilevers \cite{LoncarDia,Arcizet2011}, quantum dots on carbon nanotubes \cite{Ganzhorn2013,Palyi2012}, nanomechanical resonators coupled to quantum dots \cite{Chan2011} or superconducting qubits \cite{LaHaye2009}, and superconducting circuit QED\cite{PhysRevLett.109.240501,Pirkkalainen2013}. Each of these systems features a high quality resonator, some with extremely high -- though of course finite -- Q factors, as well as a discrete system whose interaction with the environment will in general not be entirely restricted to the resonator. 

\begin{figure}[b]
\centering
\includegraphics[scale=0.2]{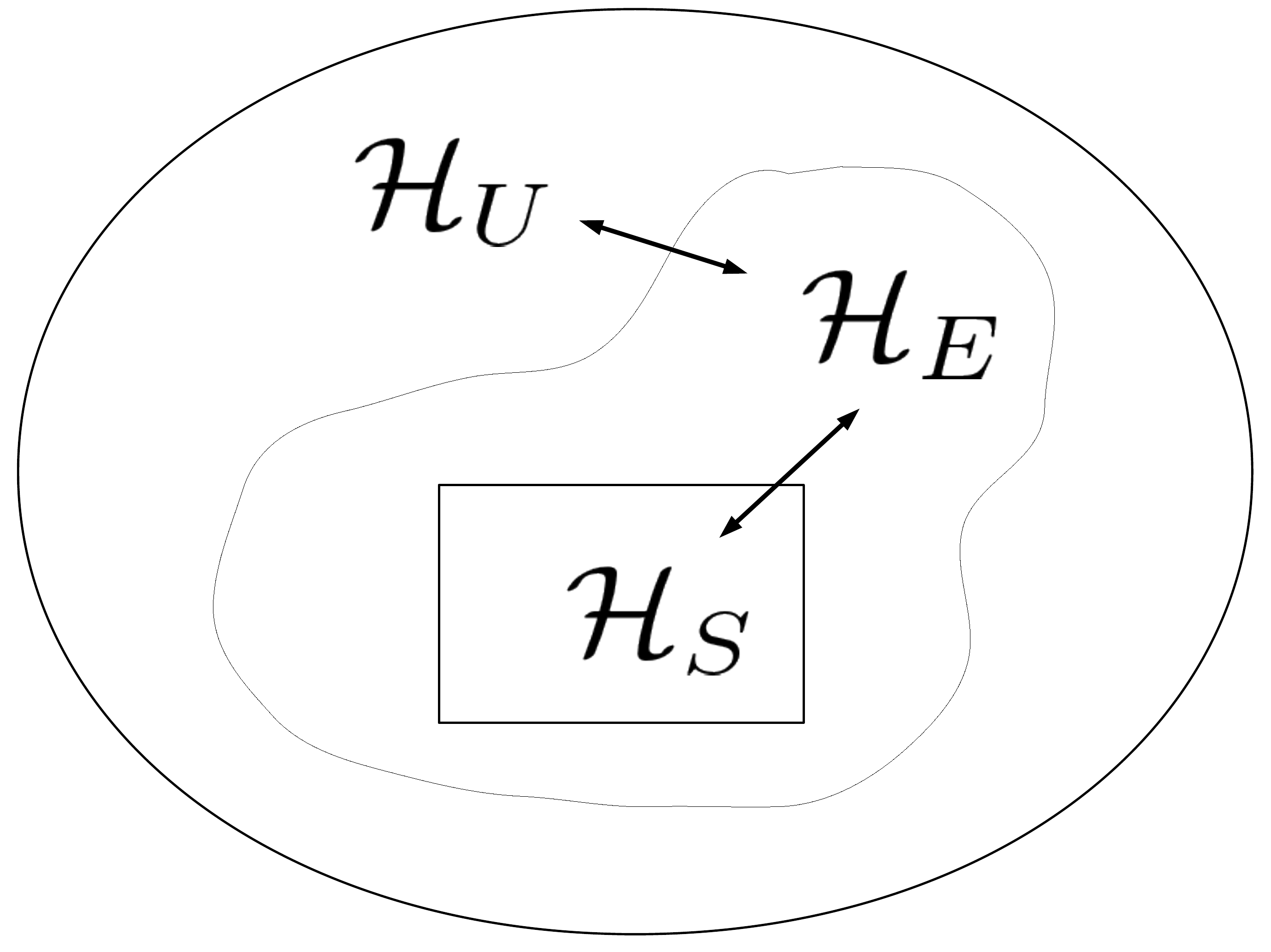}
\caption{
\label{figure illustration}
An illustration of the model under study. The system of interest is coupled to an immediate environment, which is in turn coupled to the wider `universe'. The environment is modelled as a set of harmonic oscillators, whereas the `universe' weakly dampens each of these oscillators to a thermal state.
}
\end{figure}

Additionally, our technique can be applied to the study of nanoscale energy transfer. For example, the interplay of vibrational modes and the excitonic states in molecular structures are thought to be key to fully understanding photosynthesis\cite{Engel2007}.
Indeed, a dominant coupling of an energy transfer complex to a small number of discrete vibrational modes may be responsible for efficient energy transfer\cite{O'Reilly2014}, and previous work has shown how a continuous spectrum of modes can be mapped onto a bath plus one or more coupled and discrete oscillator modes.\cite{prior2010, ilessmith2014} However, new theoretical developments, and further experiments, are needed to understand the functional role of discrete modes in energy transfer systems. The theoretical framework we describe here is ideal for studying this kind of system-discrete mode-bath system and is applicable across a wide range of parameter space. For example, it can accurately reproduce the energy transfer dynamics occurring in the FMO complex\cite{Fruchtmann2014b}.

To illustrate our method, we show that it delivers a highly accurate description of the ubiquitous Rabi model, even when the oscillator is damped by a larger environment. As a second example, we take the spin boson model, showing how our method reduces to the weak-coupling results in the appropriate limit, whilst in general giving better agreement with exact QUAPI calculations than traditional weak-coupling techniques. Moreover, since we do not restrict ourselves to the Markovian limit with a static environment, we are able to explore the case where the bath oscillators are themselves coupled to a larger environment, 
and we derive analytical expressions for the decoherence and dephasing rates for this case. 

This paper is organized as follows:  in Sec.~\ref{section model} we define our model and give a brief introduction to the coherent state representation, and introduce the influence functional. Section \ref{section a single mode} introduces the perturbative solution to the case where the environment is a single damped vibrational mode. In Sec.~\ref{section multimode} we examine the case of a more complex environment which is defined via a general spectral density, and show that up to second order in perturbation, each mode contributes independently to the dynamics. Sec.~\ref{section spin boson} studies the spin-boson model, comparing our method to other approaches, and finally, in Sec.~\ref{summary}, we summarize our results and discuss the validity of our technique.

\section{Coherent state representation and Model}\label{section model}
\subsection{Model}\label{subsection model}

We start with the Hamiltonian
\begin{gather}\label{Hamiltonian1}
\Hamiltonian = \Hamiltonian_S + \Hamiltonian_E + \Hamiltonian_{I} + \Hamiltonian_{U} + \Hamiltonian_{EU}
\end{gather}
where $\Hamiltonian_S$ is the Hamiltonian of the governing the system of interest. We shall take the ``system Hamiltonian"  to be defined on a discrete, finite-dimensional Hilbert space, on which measurements can be performed. No other assumptions are necessary, and in particular $\Hamiltonian_S$ does not need to be time-independent.
The term $\Hamiltonian_E = \sum_k \omega_k \ad_k a_k$ represents an environment consisting of harmonic oscillators, where $a_k^\dagger$ ($a_k$) is the creation (annihilation) operator for a mode with angular frequency $\omega_k$. The term $\Hamiltonian_I = V \sum_k g_k (\ad_k+a_k)$ is the interaction coupling the system (via the system operator $V$) to the environment. 

Eqn.~(\ref{Hamiltonian1}) also includes terms that allow our environment to be coupled to the rest of the universe denoted by $\Hamiltonian_U$. When such a wider environment is present, we assume that it is well approximated by an infinite heat bath that is kept in a thermal state. The oscillator modes of the immediate environment are then dynamically driven towards a thermal state by virtue of the environment to universe coupling term $\Hamiltonian_{EU}$. However, unlike conventional Born-Markov weak coupling approaches which commonly keep the entire environment fixed in thermal equilibrium, the inner tier modes will in general deviate from the thermal state. We shall show this adds an exponential cut-off to the response kernel. Figure \ref{figure illustration} gives an illustration of our model.

Instead of explicitly treating the coupling between the environment and the rest of the universe with a microscopic derivation, 
we make the simplifying assumption that $\Hamiltonian_{EU}$ is small enough that each mode $\omega_k$ of the environment simply experiences damping with rate $\gamma_k$ via standard Lindblad operators (for a derivation see, e.g., Ref.~\onlinecite{breuer2007theory}).
For this to be consistent, two conditions must be satisfied: 
Firstly, the damping rate $\gamma_k \ll \omega_k$ must be small for each mode, because this is the parameter regime assumed in the derivation of the damped harmonic oscillator master equation. Secondly, the system-environment coupling described by $\Hamiltonian_I$ may not become too large either,

otherwise the damping Lindblad operators acting on each mode are influenced by the presence of the system and our simple independent choice ceases to be a good approximation~\cite{StrongInteractions}
 (also see Ref.~\onlinecite{Scala2007} for a discussion of this approximation in the context of the resonant damped Jaynes-Cummings model).

Finally, we assume that the initial density matrix can be factorized as $\rho(0) = \rho_s(0)\otimes\rho_E^{th}$ with the initial thermal state of the environment being $\rho_E^{th} = {\cal N}^{-1}\exp(-\beta \Hamiltonian_E)$ (where ${\cal N}$ is the appropriate normalization factor).

\subsection{Coherent representation} 
To represent the density matrix of a single harmonic oscillator we use the \emph{coherent state} or \emph{P representation}\cite{gardiner2004quantum}, which has been extensively studied in quantum optics. The coherent state representation maps between the density matrix of a harmonic oscillator $\rho$ and a function of two continuous variables $P(\alpha,\alphas)$ via
\begin{gather}
\rho = \int d^2\alpha P(\alpha,\alphas)\ket{\alpha}\bra{\alpha} ~,
\end{gather}
where $\ket{\alpha}$ is the coherent state defined as $\ket{\alpha} = e^{\alpha\ad-\alphas a}\ket{0}$ or alternatively $a\ket{\alpha} = \alpha\ket{\alpha}$, and $d^2\alpha \equiv d\text{Re}(\alpha)d\text{Im}(\alpha)$. The mapping yields the following operator correspondence~\cite{gardiner2004quantum}:
\begin{align}
a \rho &\leftrightarrow \alpha P ~, \\
\rho \ad &\leftrightarrow \alphas P ~, \\
\ad \rho &\leftrightarrow (\alphas-\dalpha) P ~, \\
\rho a &\leftrightarrow (\alpha-\dalphas) P~.
\end{align}
For a system with states $\ket{i}$ coupled to an oscillator, instead of a $P$ function we now need a $P$ matrix to represent the density matrix,
\begin{gather}
\rho = \sum_{i,j}\int d^2\alpha P_{i,j}(\alpha,\alphas)\ket{i,\alpha}\bra{j,\alpha}.
\end{gather}
Generalizing from a single mode to a set of modes is straightforward, 
with the corresponding set of variables $\{a_k,\ad_k\} \leftrightarrow \{\alpha_k, \alphas_k\}$ and
\begin{gather}
\rho = \sum_{i,j}\left(\prod_k\int d^2\alpha_k\right) P_{i,j}(\{\alpha_k,\alphas_k\})\ket{i,\{\alpha_k\}}\bra{j,\{\alpha_k\}}.
\end{gather}
A partial trace over the oscillator space is given by
\begin{gather}
\text{Tr}_\text{osc}(\rho) = \sum_{i,j}\left(\prod_k \int d^2 \alpha_k\right) P_{i,j}(\{\alpha_k,\alphas_k\})\ket{i}\bra{j}.
\end{gather}
For notational ease, from hereon we switch to a vectorized form of the density matrix and operators, mapping $n\times n$ matrices $A_{i,j}$ to vectors  ${A_i}$ of dimension $n^2$.
Further, we use the generalized Gell-Mann matrices with the notation from Ref.~\onlinecite{Gell-MannMat}. For an $n$-site system, these consist of $n^2-1$ traceless and Hermitian matrices $\nu_1,\nu_2,\cdots, \nu_{n^2-1}$, defining a full operator basis together with the identity matrix.\footnote{For $n=2$ (a qubit) $\nu_i = \sigma_i$ are the Pauli matrices, and for $n=3$ we get the Gell-Mann matrices $\nu_i=\lambda_i$.}
Adopting the Einstein summation convention, where $i,j,k$ run from $1$ to $n^2-1$, the generalized Gell-Mann matrices satisfy:
\begin{gather}
\nu_i \nu_j = \frac{2}{n}\delta_{i j}+(d_{ijk}+i f_{ijk})\nu_k \\
[\nu_i, \nu_j] = 2 i f_{ijk} \nu_k \\
\{\nu_i, \nu_j\} = \frac{4}{n}\delta_{ij}+2 d_{ijk} \nu_k,
\end{gather}
where $f_{ijk}$ and $d_{ijk}$ are totally antisymmetric and symmetric tensors, respectively.
For $n=2, f_{ijk} = \epsilon_{ijk}$ the Levi-Civita symbol and $d_{ijk} = 0$. Any $n\times n$ matrix $P$ can be written as a vector $P_i$:
\begin{gather}
P = P_{n^2}\id + P_i \nu_i ~, \\
P_{i} = \half Tr[P \nu_i] ~, \\
P_{n^2} = (1/n) Tr[P] ~.
\end{gather}
Using this vectorized form we can write the density matrix as
\begin{gather}\label{general den matrix}
\rho = \int_\alpha \Big(P_{n^2}\id+P_i \nu_i\Big) \ket{\{\alpha_k\}}\bra{\{\alpha_k\}} ~,
\end{gather}
where for convenience we denote $\int_\alpha \equiv \prod_k\int d^2 \alpha_k$, and $P=P(\{\alpha_k,\alphas_k\})$.
The condition $Tr\rho = 1$ implies $\int d^2\alpha P_{n^2}(\alpha,\alphas) = 1/n$, and we are interested in the partial trace over the environment
\begin{gather}
\rho_s = \int_\alpha \left(P_{n^2}\id+P_i \nu_i \right) \equiv (1/n)\id+\rho^s_i \nu_i ~.
\end{gather}

\subsection{The Influence Functional}
At this stage, we use the following form for writing down the full dynamics of the reduced system:
\begin{gather}\label{ansatz1}
\rho^s(t) = U(t)e^{\Theta(t)}\rho^s(0) ~,
\end{gather}
where $U(t)$ is the propagator (in the vectorized representation) of the system without the environment, and the influence of the rest of the world on the system is encoded in the influence functional $\Theta(t)$. 
The motivation for this comes from the Feynman-Vernon influence functional~\cite{Feynman1963} of the same form. Further, we anticipate that this form will be a convenient one for recovering the known exponential decay in the weak-coupling limit. 
The main result of this paper is that it is possible to find an exact expansion of $\Theta(t)$ as a perturbation series with respect to the interaction $\Hamiltonian_I$, and expansion up to second order recovers the known dephasing and relaxation rates given by standard Born-Markov weak master-equation techniques, but with an added non-Markovian contribution.

\section{A Single Mode}\label{section a single mode}
Let us first examine the case where the environment $\Hamiltonian_E = \omega \ad a$ consists of only a single mode. When taking a two-level system (2LS) as the system (a limitation which is not required in the following), then this is just the well-known Rabi model.

In its vectorized form, the system-environment part of Hamiltonian (\ref{Hamiltonian1}) can be decomposed to
\begin{gather}\label{single Mode Hamiltonian 1}
\Hamiltonian_S(t) = H_i(t) \nu_i~,\\
\Hamiltonian_E = \omega \ad a~,\\
\Hamiltonian_I(t) = g V(t) (a+\ad)~,\\
V(t) = V_i(t) \nu_i +V_{n^2}(t)\id~.\label{single Mode Hamiltonian 2}
\end{gather}
Then the operator correspondence between $\rho$ and $\vec{P}$, with the vector $\vec{P} = [P_1(\alpha),P_2(\alpha),\cdots,P_{n^2}(\alpha)]$ yields:
\begin{gather}
\pd{}{t}\rho = -i[\Hamiltonian_S + \Hamiltonian_E + \Hamiltonian_I,\rho] + D(\rho) \leftrightarrow \notag\\
\pd{}{t}\vec{P} = -i(\Hamiltonian_S^\times+L)\vec{P} + g A_g \vec{P}\label{equation total equivalence} ~.
\end{gather}
Here $D(\rho)$ is the Lindblad dissipator induced by $\Hamiltonian_U + \Hamiltonian_{EU}$, which damps the oscillator with rate $\gamma$. The operator 
\begin{align}
L =& \notag(-\omega+\frac{i}{2}\gamma)\dalpha\alpha + (\omega+\frac{i}{2}\gamma)\dalphas\alphas \\ &+ i\gamma N  \pd{^2}{\alpha\partial \alphas}
\end{align}
is simply the corresponding P representation Fokker-Plank operator\citep{breuer2007theory}, i.e.~for a single damped oscillator the Master Equation would read $\pd{}{t}P = -i L P$, where $N=[\exp(\beta \omega)-1]^{-1}$ is the mean oscillator occupation number at thermal equilibrium with inverse temperature $\beta = (k_b T)^{-1}$. In the vectorized representation, the terms $-i\Hamiltonian_S^\times P$ and $g A_gP$  take the place of $-i[\Hamiltonian_S,\rho]$ and $-i[\Hamiltonian_I,\rho]$, respectively, where the matrices $\Hamiltonian_S^\times, A_g$ are given by
\begin{gather}
\left[\Hamiltonian_S^\times(t)\right]_{ij} = -2 i H_k(t) f_{kij}~,\\
\left(\Hamiltonian_S^\times \right)_{i, n^2} = \left(\Hamiltonian_S^\times \right)_{n^2, i} = 0~,\\
\left[A_g(t)\right]_{ij} = -i\left(\pd{}{\alphas}-\pd{}{\alpha}\right)[V_k(t) d_{kij}+V_{n^2}(t)\delta_{ij}]\notag\\ \;\;\;\;\;\;\;-\left(2\alpha+2\alphas-\pd{}{\alpha}-\pd{}{\alphas}\right)V_k(t) f_{kij}~,\\
\left[A_g\right]_{i, n^2} = -i\left(\pd{}{\alphas}-\pd{}{\alpha}\right)V_i~,\\
\left[A_g\right]_{n^2, i} = -i\left(\pd{}{\alphas}-\pd{}{\alpha}\right)\frac{2}{n}V_i(t)~,\\
\left[A_g\right]_{n^2, n^2} = -i\left(\pd{}{\alphas}-\pd{}{\alpha}\right)V_{n^2}(t)~.
\end{gather}
Note that $\Hamiltonian_S^\times$ is Hermitian, and the propagator $U(t)$ satisfies
\begin{gather}
\pd{}{t}U(t) = -i \Hamiltonian_S^\times U(t)~,\\
U(0) = \id~.
\end{gather}

The central strategy of this paper now is to solve Eqn.~(\ref{equation total equivalence}) perturbatively with $g$ being the small parameter, based on the form (\ref{ansatz1}) of the full solution in order to estimate the influence functional $\Theta(t)$.

\subsection{Perturbation Series}
For the perturbation treatment, we use the expansion
\begin{gather}\label{pertur nsites}
P = P^0 + g P^1 + g^2 P^2 + \cdots ~,
\end{gather}
hence Eqn.~(\ref{equation total equivalence}) translates to:
\begin{align}
\pd{}{t} P^0 &= -i(\Hamiltonian_S^\times+L) P^0 \label{P0 eq nsites} ~, \\
\pd{}{t} P^1 &= -i(\Hamiltonian_S^\times+L) P^1 + A_g P^0 \label{P1 eq nsites} ~, \\
\pd{}{t} P^2 &= -i(\Hamiltonian_S^\times+L) P^2 + A_g P^1 ~, \\
\cdots\notag
\\
\pd{}{t} P^n &= -i(\Hamiltonian_S^\times+L) P^n + A_g P^{n-1} \label{Pn eq nsites} ~.
\end{align}
The solution for the uncoupled system $P^0$ is simply given by
\begin{gather}
P^0(t) = U(t)\rho^s(0)
\frac{1}{\pi N}e^{-|\alpha|^2/N}
\end{gather}
with
$\rho^s(t) = [\rho^s_1(t) , \rho^s_2(t) , \dots , \rho^s_{n^2-1}(t) , 1/n]$.
In principle it is possible to solve this series term by term. However, we are interested in the state of the system and not the oscillator, which makes things much easier: We use the boundary condition where $\alpha^k P^n(\alpha)\underset{\alpha\rightarrow\infty}\longrightarrow 0$ for all $k,n$. This is justified since the oscillator can be expected not to deviate by too much from a thermal, Gaussian state, and it certainly also should not occupy extreme high-energy states. Therefore performing the integration $\int d^2\alpha\equiv\int_{\alpha}$ on Eqn.~(\ref{P1 eq nsites}-\ref{Pn eq nsites}) yields 
\begin{gather}
\pd{}{t} \int_\alpha P^1 = -i \Hamiltonian_S^\times \int_\alpha P^1 -i V^\times \underbrace{\int_\alpha (\alpha+\alphas)P^0}_{\rightarrow 0} \label{P1 inteq} ~, \\
\pd{}{t} \int_\alpha P^2 = -i \Hamiltonian_S^\times \int_\alpha P^2 -i V^\times\int_\alpha (\alpha+\alphas)P^1 \label{P2 inteq nsites} ~, \\
\cdots\notag
\\
\pd{}{t} \int_\alpha P^n = -i \Hamiltonian_S^\times \int_\alpha P^n -i V^\times\int_\alpha (\alpha+\alphas)P^{n-1} \label{Pn inteq nsites} ~, 
\end{gather}
where
\begin{gather}\label{A_1 def}
\left(V^\times\right)_{ij} = -2 i V_k f_{kij} ~,
\\
\left(V^\times\right)_{i,n^2} = \left(V^\times\right)_{n^2,i} = \left(V^\times\right)_{n^2,n^2} = 0 ~,
\end{gather}
is the matrix equivalent to the superoperator $[V,\square]$.
The initial condition is 
$\int_\alpha P^{n>0}(t=0) = 0$, i.e.~at time $t=0$ the qubit and the mode are factorized, and the mode is in the thermal state, which gives 
\begin{gather}
\int_\alpha P^1(\alpha,t) = 0
\end{gather}
for all times. The first contribution in the expansion therefore comes from $\int_\alpha P^2(\alpha,t)\neq 0$, which is $2^{nd}$ order in the coupling constant $g$. This is in analogy to the usual QME treatment, where the influence of the environment also enters at the $2^{nd}$ order in the coupling constant.
In order to solve Eqn.~(\ref{P2 inteq nsites}) we first need to evaluate $\int_{\alpha} (\alpha+\alphas) P^1$, which can be done by invoking the following mathematical procedure: 
(i) multiply Eqn.~(\ref{P1 eq nsites}) by $\alpha$ or $\alphas$ from the left; (ii) perform the $\int_\alpha$ integral; (iii) integrate by parts all terms possessing a derivative. The sequence of these steps  yields the following two equations:
\begin{gather}\label{originaltwoeq gen}
\left[ \pd{}{t}+i\omega+\half\gamma +i\Hamiltonian_S^\times(t) \right] \int_\alpha \alpha P^1 = \int_{\alpha} \alpha A_g (t)P^0 ~,\\
\left[ \pd{}{t}-i\omega+\half\gamma +i\Hamiltonian_S^\times(t) \right] \int_\alpha \alphas P^1 = \int_{\alpha} \alphas A_g (t)P^0 ~, \label{originaltwoeq gen2}
\end{gather}
which after a bit of algebra and ODE solving yield a solution for $\int_\alpha P^1$. Substituting this solution into Eqn.~(\ref{P2 inteq nsites}) then results in
\begin{gather}\label{intalphaP2}
\int_\alpha P^2 = - U(t) \int_0^t dt' \int_0^{t'} dt'' e^{-\half \gamma(t'-t'')}\tilde{V}^\times(t')\times
\\
\Big[
(2N+1)\cos[\omega(t'-t'')] \tilde{V}^\times(t'') \notag\\
-i\sin[\omega(t'-t'')]\tilde{V}^\circ(t'')\Big] \rho^s(0)\notag ~.
\end{gather}
Here, the notation $\tilde{V}^\times,\tilde{V}^\circ$ denotes operators in the Heisenberg picture,
\begin{gather}\label{a1a2 def}
\tilde{V}(t) \equiv U^{-1}(t)V(t) U(t) ~,
\end{gather}
and $\tilde{V}^\circ$ is the equivalent of $\{V,\Box\}$ and is given by
\begin{gather}
\left(V^\circ\right)_{i, n^2} = 2 V_i(t) ~,\\
\left(V^\circ\right)_{n^2, i} = \frac{4}{n} V_i(t) ~,\\
\left(V^\circ\right)_{i j} = 2 V_k(t) d_{kij}+2V_{n^2}(t)\delta_{i,j} ~, \\
\left(V^\circ\right)_{n^2 n^2} = 2 V_{n^2}(t) ~.
\end{gather}

At this point we note that the influence functional $\Theta(t)$ up to second-order in $g$ is then given by Eqn.~(\ref{intalphaP2}) and
\begin{gather}\label{Theta equation single}
U(t)\Theta(t)\rho^s(0) = g^2\int_\alpha P^2.
\end{gather}
We proceed by showing that this provides a highly accurate solution for the single mode case in the weak-coupling limit. We shall then generalise the technique to an environment consisting of a (quasi)continuous bath of oscillators.
In Appendix \ref{appendix higherOrders} we sketch the derivation of higher-order terms in the perturbation series.

\subsection{Example: the (damped) Rabi model}

\begin{figure}
\centering
\includegraphics[scale=0.8]{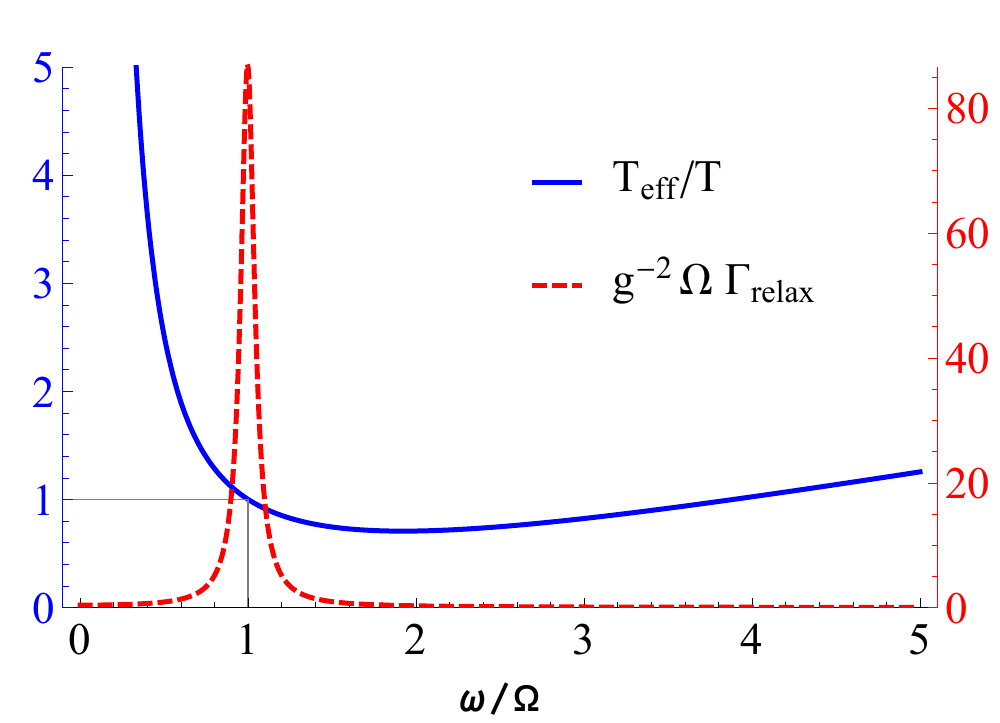}
\caption{
\label{figure Teff}
The apparent effective temperature  of the system as defined by  
Eqn.~(\ref{effTemp eq}) (blue), and the relaxation constant $\Omega\Gamma_\text{relax} / g^2$, as in Eqn.~\ref{relaxation constant eq}, (dashed red) as a function of $\omega / \Omega$. Other parameters are: $\beta \Omega = 1$, $\gamma/\omega = 0.1$ and $\epsilon = 0$ (no bias).
}
\end{figure}

The Rabi model, consisting of a coupled 2LS to a harmonic oscillator, represents perhaps the most basic and ubiquitous compound quantum system. Focussing only on the dynamics of the 2LS and tracing over the oscillator then results in arguably the conceptually most simple and yet a highly non-trivial open systems problem. Let us consider the Rabi Hamiltonian
\begin{gather}\label{Rabi Hamiltonian}
\Hamiltonian = \frac{\epsilon}{2} \sigma_z + \frac{\Delta}{2} \sigma_x +\omega\ad a + g(a+\ad)\sigma_z + \Hamiltonian_{EU} + \Hamiltonian_{U}~,
\end{gather}
where $\sigma_i$ are the usual Pauli matrices referring to the 2LS. In this case, we immediately find that the matrices $\Hamiltonian_S^\times, V^\times, V^\circ$ are given by:
\begin{gather}
\label{A_0 Rabi Model}
\Hamiltonian_S^\times \equiv \begin{pmatrix}
0 & -i\epsilon & 0 & 0 \\
i\epsilon & 0 & -i\Delta & 0\\
0 & i\Delta & 0 & 0\\
0 & 0 & 0 & 0
\end{pmatrix} ~, \\
V^\times = \begin{pmatrix}
0 & -2i & 0 & 0 \\
2i & 0 & 0 & 0\\
0 & 0 & 0 & 0\\
0 & 0 & 0 & 0
\end{pmatrix}
\label{A_1 Rabi Model} ~,
\\
V^\circ = \begin{pmatrix}
0 & 0 & 0 & 0 \\
0 & 0 & 0 & 0\\
0 & 0 & 0 & 2\\
0 & 0 & 2 & 0
\end{pmatrix}  
\label{A_2 Rabi Model}
~,
\end{gather}
when operating on the vector $\{\sigma_x, \sigma_y, \sigma_z, \id\}^\dagger$.
Substituting these into Eqn.~(\ref{Theta equation single}), we obtain an unwieldy analytical expression for $\Theta(t)$, which can give us insight if examined in the eigenbasis of the system (the $\Hamiltonian_S^\times$ eigenbasis):
the top $3\times 3$ part of $\Hamiltonian_S^\times$ has two finite and one vanishing eigenvalue ($\{0, \pm\sqrt{\epsilon^2+\Delta^2}\}$). 
In this basis, the real terms on the diagonal of $\Theta(t)$ that are proportional to $t$ and correspond to the finite eigenvalues, are both equal to the dephasing rate. The one corresponding to the vanishing eigenvalue is the relaxation rate. These rates are given by
\begin{align}
\Gamma_\text{relax} =&
\label{relaxation constant eq}
\\
g^2 \coth(\frac{\beta \omega}{2}) &\frac{\Delta^2}{\Omega^2}
\left(\frac{\gamma}{(\frac{\gamma}{2})^2+(\Omega-\omega)^2}+\frac{\gamma}{(\frac{\gamma}{2})^2+(\Omega+\omega)^2}\right)
\notag ~,
\\
\label{dephase constant eq}
\Gamma_\text{dephase} = \frac{1}{2}&\Gamma_\text{relax}+ 
2 g^2 \coth(\frac{\beta \omega}{2}) \frac{\epsilon^2}{\Omega^2}
\frac{\gamma}{(\frac{\gamma}{2})^2+\omega^2} ~,
\end{align}
where $\Omega = \sqrt{\epsilon^2+\omega^2}$ is the Rabi frequency. Note that in the limit $\gamma\rightarrow 0$, i.e.~no damping on the oscillator from the wider environment or universe, we recover the standard Born-Markov ME result for relaxation and dephasing, given in Eqns.~(\ref{eq appendix relaxation}-\ref{eq appendix dephasing}).
 The imaginary parts on the diagonal of $\Theta(t)$ correspond to the Lamb shift Hamiltonian, given by
\begin{align}\label{lamb shift eq}
\Hamiltonian_{LS} &= \frac{1}{2}\tilde{\sigma}_z g^2 \coth(\frac{\beta \omega}{2}) \frac{\Delta^2}{\Omega^2} \times \\ 
&\left(\frac{\Omega-\omega}{(\frac{\gamma}{2})^2+(\Omega-\omega)^2}+\frac{\Omega+\omega}{(\frac{\gamma}{2})^2+(\Omega+\omega)^2}\right),
\notag
\end{align}
where $\tilde{\sigma_z}$ is given by writing the system Hamiltonian, i.e. the first two terms in Eqn.~(\ref{Rabi Hamiltonian}) in its diagonal basis 
\begin{gather}
\tilde\Hamiltonian_S = \half \Omega \tilde{\sigma}_z~.
\end{gather}
Again, in the limit $\gamma\rightarrow 0$ we recover the ``standard'' Lamb shift given in Eqn.~(\ref{eq appendix lambshift}).
Furthermore, we can extract the steady state of the system at long times: 
At times much larger than the relaxation time, the system tends to the state
\begin{align}\label{effTemp eq}
\rho(t\gg &\Gamma_\text{relax}^{-1} )\rightarrow 
\\
&\half - \half\tilde{\sigma}_z \frac{2\Omega\omega}{(\frac{\gamma}{2})^2+\Omega^2+\omega^2}\tanh(\frac{\beta \omega}{2})~.\notag
\end{align}
This is indeed only the expected thermal system state when $\gamma\rightarrow 0$ and $\omega\rightarrow\Omega$, i.e.~no damping and when oscillator and system are resonant. However, one should take this limit with caution, because for vanishing damping, $\gamma\rightarrow 0$ the relaxation time $\Gamma_\text{relax}^{-1}$ tends to infinity 
 and the system will thus never actually reach this state.
In Fig.~\ref{figure Teff} we plot the effective temperature, that is, the temperature $T_\text{eff}$ given by equating $\exp[-\tilde{\Hamiltonian}_S / k_b T_\text{eff}]$ with Eqn.~(\ref{effTemp eq}). On the same figure we plot the relaxation rate for the same parameters, showing a Lorentzian peak in efficiency near resonance.

We note that in general the effective temperature differs from the temperature of the universe.
In order to explain this apparent discrepancy, we examine Eqn.~(\ref{effTemp eq}): The universe is only directly coupled to the oscillator which has energy levels spacing of $\omega$, this accounts for the term $\tanh(\frac{\beta \omega}{2})$ which is different from the expected $\tanh(\frac{\beta \Omega}{2})$. This term decreases (increases) the effective temperature $T_\text{eff}$ when the mode is blue-shifted (red-shifted) with respect to the Rabi frequency $\Omega$.
The pre-factor
\begin{gather}
\frac{2\Omega\omega}{(\frac{\gamma}{2})^2+\Omega^2+\omega^2} = 1 - \frac{(\Omega-\omega)^2+(\frac{\gamma}{2})^2}{(\frac{\gamma}{2})^2+\Omega^2+\omega^2}
\end{gather}
is maximized when on resonance ($\omega=\Omega$). Detuning suggests that in order to extract energy from the qubit, the universe exchanges energy with the oscillator to match the detuning. This adds uncertainty to the system effectively increasing the temperature. The system-environment coupling $\gamma$ adds additional uncertainty.

We also note that in this scheme we do not keep track of the environment, only trace over it. The thermal state of system+environment is proportional to $\exp[-\beta (\Hamiltonian_S+\Hamiltonian_E+\Hamiltonian_I)]$, i.e. the system and environment are entangled, and defining a temperature of just one subsystem is questionable.

The example we discuss in this section is formally equivalent to the reaction coordinate~\cite{Garg1985,Thorwart2000,Hughes2009} or structured environment~\cite{Thorwart2004,Goorden2004,Goorden2005,Huang2008} model in the weak coupling and weak damping regime. Here, the reaction coordinate model employs an effective spectral density with a Lorenzian peak, yielding the same rates as Eqns.~(\ref{relaxation constant eq}-\ref{lamb shift eq}) except for the ``counter rotating'' terms $\sim (\Omega+\omega)^{-n}$ (which are typically small). Interestingly however, this nice agreement only extends to the real part of the response function, $D(t)$, which determines the damping rates. By contrast, the modified spectral density of the reaction coordinate method does not account for corrections to the imaginary part $D_1(t)$, which yields the long time asymptotic behaviour of the system. 
To ensure that our approach does indeed deliver the correct steady state, we have made a comparison with an exact numerical simulation of the dynamics given by Hamiltonian (\ref{Rabi Hamiltonian}) (with $\Hamiltonian_{EU} + \Hamiltonian_{U}$ replaced by a Lindblad dissipator). We obtain perfect agreement between Eqn.~(\ref{effTemp eq}) and a purely numerical simulation in the weak coupling regime.

\begin{figure}
\centering
\includegraphics[scale=0.45]{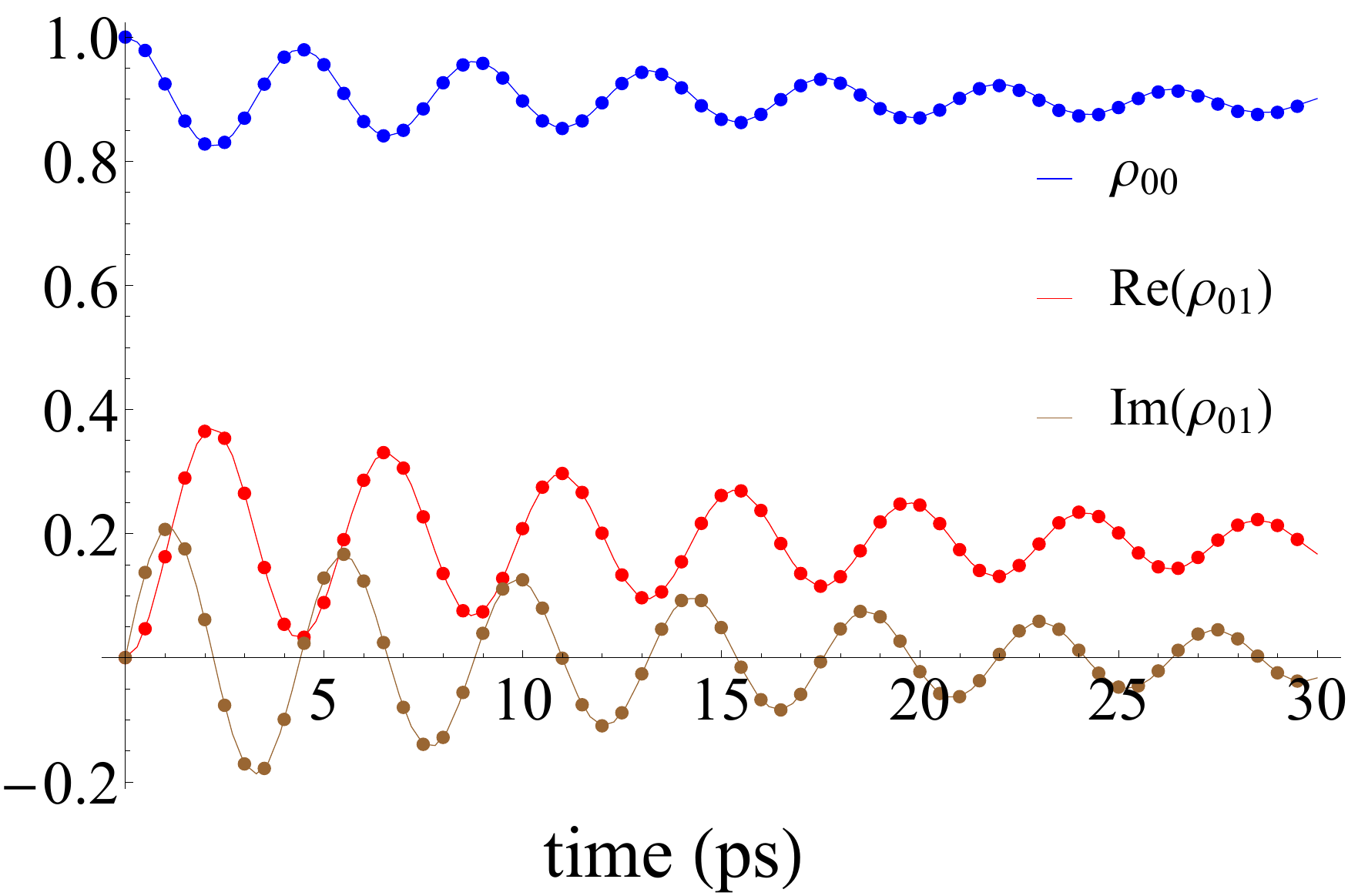}
\caption{
\label{figure Rabi Model}
A comparison between the dynamics given by Eqn.~(\ref{ansatz1}) with $\Theta(t)$ approximated by Eqn.~(\ref{Theta equation single}) (solid) and exact numerical simulation of Hamiltonian (\ref{Rabi Hamiltonian}) dynamics (dotted). 
The Parameters used here are
$\Delta = 0.6 \text{ ps}^{-1}$ , 
$\gamma = 0.8 \text{ ps}^{-1}$ ,
$\epsilon = 1.3 \text{ ps}^{-1}$ ,
$\omega = 0.2 \text{ ps}^{-1}$ ,
$k_b T = 1 \text{ ps}^{-1}$,
$g = 0.03$.
The approach to equilibrium is not prominent in this case because of the long relaxation time $\Gamma_\text{relax}^{-1}\approx 3000\text{ps}$. 
The dephasing time is much shorter with $\Gamma^{-1}_\text{dephase}\approx 17\text{ ps}$.}
\end{figure}

In Figure \ref{figure Rabi Model} we plot a comparison between Eqn.~(\ref{ansatz1}) with $\Theta(t)$ approximated by Eqn.~(\ref{Theta equation single}), and exact numerical simulation, showing that for the weak-coupling regime there is a very good agreement between the two.

\section{Extending the analysis to a multimode environment}\label{section multimode}
In the previous section the `environment' consisted of only one single harmonic oscillator. However, adding multiple oscillators is straightforward, and in the weak coupling limit, where environmental influence is assumed to be small, each environmental mode contributes to the influence functional $\Theta(t)$ independently.
The difference is that now the environment Hamiltonian $\Hamiltonian_E$ has a set of modes, and in our vectorized form the equivalent of Eqns.~(\ref{single Mode Hamiltonian 1}-\ref{single Mode Hamiltonian 2}) becomes
\begin{gather}\label{multi mode Hamiltonian 1}
\Hamiltonian_S(t) = H_i(t) \nu_i ~,\\
\Hamiltonian_E = \sum_k \omega_k \ad_k a_k ~,\\
\Hamiltonian_I(t) = \sum_k g_k V(t) (a_k+\ad_k) ~, \\
V(t) = V_i(t) \nu_i + V_{n^2}\id ~.\label{multi mode Hamiltonian 2}
\end{gather}

The derivation for this case is very similar to the single mode case and is given in full detail in Appendix \ref{appendix multimode}. Once more,  
the influence of the bath on the system's dynamics is given by Eqn.~(\ref{ansatz1}), where now
\begin{align}\label{eq Theta multimode}
\Theta(t) = &-\int_0^t dt' \int_0^{t'}dt'' \tilde{V}^\times(t')\times
\notag\\
&\left[D_\gamma(t'-t'')\tilde{V}^\times(t'')+i D_{\gamma 1}(t'-t'')\tilde{V}^\circ(t'')\right].
\end{align}
Here $\tilde{A}_{1,2}$ are given by Eqn.~(\ref{a1a2 def}), and we adapt our notation to match that common in the literature on phonon baths, introducing the (damped) phonon response function defined as
\begin{align}
\alpha_\gamma(\tau) &= \sum_k g_k^2 e^{-\half \gamma_k \tau}\frac{\cosh(\frac{\beta \omega_k}{2}-i\omega_k \tau)}{\sinh{(\frac{\beta \omega_k}{2})}} 
\notag\\
&\equiv D_{\gamma}(\tau) + i D_{1\gamma}(\tau)~.
\label{dampedKernel}
\end{align}
Here $D_\gamma(\tau)$ and $D_{1\gamma}(\tau)$ are the (damped) dissipation and response kernels, respectively.
In terms of the spectral density function,
\begin{gather}\label{eq Spectral density}
J(\omega) = \sum_k g_k^2 \delta(\omega-\omega_k) ~,
\end{gather}
we can express the response function as
\begin{align}
\alpha_\gamma(\tau) &= \int_0^\infty d\omega e^{-\half \gamma(\omega)\tau}J(\omega)\frac{\cosh(\frac{\beta \omega}{2}-i\omega \tau)}{\sinh{(\frac{\beta \omega}{2})}} ~,
\end{align}
where $\gamma(\omega)$ is the damping rate of modes with angular frequency $\omega$. If the modes are not damped, i.e.~for $\gamma(\omega)=0$, we recover the standard response function from the literature~\cite{breuer2007theory} $\alpha(\tau) = D(\tau)+iD_1(\tau)$.

We note that for the case of $\gamma(\omega)=0$, i.e. when there is no external universe, the result (\ref{eq Theta multimode}) is exactly coincides with the well-studied time-convolutionless projection operator technique (TCL) from the literature when the TCL generator is expanded to second order in the system-environment coupling, cf. Ref.~\onlinecite{Breuer2004}.

It is interesting to note that the thermalisation of the immediate environment by the wider universe is fully captured by switching to the above generalised form of the response kernel (\ref{dampedKernel}) (within a perturbative treatment to second order, higher orders give additional corrections, see Appendix \ref{appendix higherOrders}). At $T=0$ our expression is in full agreement with the previously derived zero temperature response function of the damped spin-boson model given in Ref.~\onlinecite{Imamog1994}. We suggest that the same kernel redefinition might also be applicable to other methods of studying open quantum systems, giving a simple recipe to adding a wider universe on top of a standard open system. 

\subsection{Example: The Spin-Boson Model}\label{section spin boson}
To apply our generalized multimode technique to a particular example, we look at the well studied case of the (biased) spin-boson model with the following Hamiltonian:
\begin{gather}\label{eq Spin-Boson Hamiltonian}
\Hamiltonian_{SE} = \half \epsilon \sigma_z + \half \Delta \sigma_x + \sum_k \omega_k \ad_k a_k + \sigma_z\sum_k g_k (a_k+\ad_k) ~.
\end{gather}
In this case, just like for the Rabi model, the system is two-dimensional and its P vector has 4 components ($\sigma_x, \sigma_y,\sigma_z, \id$), and $\Hamiltonian_S^\times, V^\times, V^\circ$ are again given by Eqns.~(\ref{A_0 Rabi Model}-\ref{A_2 Rabi Model}).
Since we have already calculated the relaxation and dephasing rates for the single mode case, showing that the different modes contribute independently for $\Theta(t)$ in the weak-coupling regime, we can immediately write down the following expressions for the relaxation rates: we only need to add a summation $\sum_k$ over the different modes to Eqns.~(\ref{relaxation constant eq}-\ref{dephase constant eq}):
\begin{align}
\Gamma_\text{relax} =
\label{relaxation constant eq multimode}
\sum_k g_k^2 &\coth(\frac{\beta \omega_k}{2}) \frac{\Delta^2}{\Omega^2}\times
\\
\Big(&\frac{\gamma_k}{(\frac{\gamma_k}{2})^2+(\Omega-\omega_k)^2}+\frac{\gamma_k}{(\frac{\gamma_k}{2})^2+(\Omega+\omega_k)^2}\Big)~,&
\notag
\\
\label{dephase constant eq multimode}
\Gamma_\text{dephase} = \frac{1}{2}&\Gamma_\text{relax}+ 
2 \sum_k g_k^2 \coth(\frac{\beta \omega_k}{2}) \frac{\epsilon^2}{\Omega^2}
\frac{\gamma_k}{(\frac{\gamma_k}{2})^2+\omega_k^2}~.
\end{align}
We note that, as discussed at the end of Section \ref{section multimode}, in the limit of $\gamma_k\rightarrow 0$, we recover the known weak-coupling rates, cf. Ref.~\onlinecite{weiss2012quantum} or Appendix \ref{appendix weakCoupling}. The second part of Eqn.~(\ref{dephase constant eq multimode}) is known as the pure dephasing constant.

Below we study the no-bias case, setting $\epsilon = 0$: the system Hamiltonian ($\Hamiltonian_S^\times$ in our language) is static, hence the propagator $U$ is given by $U = \exp[-i \Hamiltonian_S^\times t]$.
To calculate $\Theta(t)$, we can make a change of variables in the double integral $\int_0^t dt' \int_0^{t'} dt'' = \int_0^t d\tau \int_{\tau/2}^{t-\tau/2}d\eta
$
to get the expression:
\begin{gather}\label{eq theta for spin-boson}
\Theta(t) = \Theta_\text{relax}(t)+\Theta_\text{LS}(t)+\Theta_\text{th}(t)+\Theta_\text{RW}(t)
\end{gather}
with
\begin{gather}
\Theta_\text{relax} = -2\int_0^t d\tau D_\gamma(\tau) (t-\tau)\cos\Delta\tau
\begin{pmatrix}\label{eq theta for spin-boson 1}
2 & 0 & 0 & 0\\ 0 & 1 & 0 & 0 \\ 0 & 0 & 1 & 0 \\ 0 & 0 & 0 & 0
\end{pmatrix} ~,
\\
\Theta_\text{LS} = -2\int_0^t d\tau D_\gamma(\tau) (t-\tau)\sin\Delta\tau
\begin{pmatrix}\label{eq theta for spin-boson 2}
0 & 0 & 0 & 0\\ 0 & 0 & 1 & 0 \\ 0 & -1 & 0 & 0 \\ 0 & 0 & 0 & 0
\end{pmatrix} ~,
\\ 
\Theta_\text{th} = 4\int_0^t d\tau D_{1\gamma}(\tau) (t-\tau)\sin\Delta\tau
\begin{pmatrix}\label{eq theta for spin-boson 4}
0 & 0 & 0 & 1\\ 0 & 0 & 0 & 0 \\ 0 & 0 & 0 & 0 \\ 0 & 0 & 0 & 0
\end{pmatrix} ~,
\end{gather}
\begin{align}
\label{eq theta for spin-boson 3}
\Theta_\text{RW} =-2\int_0^t d\tau D_\gamma(\tau) &\frac{1}{\Delta}\sin\Delta(t-\tau)
\times  \\
&\begin{pmatrix}
0 & 0 & 0 & 0\\ 0 & \cos\Delta t & -\sin\Delta t & 0 \\ 0 & -\sin\Delta t & -\cos\Delta t & 0 \\ 0 & 0 & 0 & 0
\end{pmatrix}~.\notag
\end{align}
In the above expression, $\Theta_\text{relax}$ induces the relaxation and decoherence, $\Theta_\text{LS}$ induces the Lamb-shift, and $\Theta_\text{th}$ steers the system towards the thermal state. $\Theta_\text{RW}$ is usually ignored under the rotating wave approximation. If one is interested in times $t \gg \tau_b$ much longer than the memory of the bath $D(t>\tau_b)\rightarrow 0$, it is justified to let the upper limit of the integrals go to infinity. For this case it is most insightful to examine this result in light of the standard quantum-optical master equation approach: In the standard approach, remarkably one gets exactly the same expressions as the above Eqn.~(\ref{eq theta for spin-boson}) [without Eqn.~(\ref{eq theta for spin-boson 3})], but with an interesting change: 
\begin{gather}
t-\tau \rightarrow t~.
\end{gather}
The terms which are \emph{not} proportional to $t$ capture non-Markovian contributions, giving information about the bath's \emph{reorganization time}. Interestingly, each of the environmental effects possesses its own timescale, and these are estimated by
\begin{gather}\label{equation reorg1}
t^R_\text{relax} = \frac{\int_0^\infty \tau d\tau D_\gamma(\tau)\cos\Delta\tau}{\int_0^\infty d\tau D_\gamma(\tau)\cos\Delta\tau} ~,
\\
t^R_\text{LS} = \frac{\int_0^\infty \tau d\tau D_\gamma(\tau)\sin\Delta\tau}{\int_0^\infty d\tau D_\gamma(\tau)\sin\Delta\tau} ~,
\\
t^R_\text{th} = \frac{\int_0^\infty \tau d\tau D_{1 \gamma}(\tau)\sin\Delta\tau}{\int_0^\infty d\tau D_{1 \gamma}(\tau)\sin\Delta\tau}\label{equation reorg2} ~.
\end{gather}
It is noteworthy that the reorganization times can be negative. 
This could happen when, for example, initially for $t \lesssim \tau_b$ the dephasing process, which includes a non-Markovian component, is more aggressive than at later times when it assumes a stable value. 
Then, as the aggressive decay stops, the population of the system has fallen by a greater amount than it would have done under the stable, long lived decay process. Thus the system appears as if it has been evolving under the stable dephasing rate for a longer time than it actually has, and hence the negative reorganization time.
We note that the terms (\ref{equation reorg1}-\ref{equation reorg2}) in the limit $\gamma\rightarrow 0$ are known in the literature as those leading to the slippage of initial conditions, and are important for preserving the positivity of the reduced density matrix.\cite{Suarez1992,Gaspard1999}

The steady-state of the system is given by
\begin{align}
\rho(t\gg &\Gamma_\text{relax}^{-1} )\rightarrow 
\\
&\half + \half \sigma_x \frac{\int_0^\infty d\tau D_{1\gamma}(\tau)\sin\Delta\tau}{\int_0^\infty d\tau D_\gamma(\tau)\cos\Delta\tau}~.\notag
\end{align}

\begin{figure}
\centering
\includegraphics[scale=0.8]{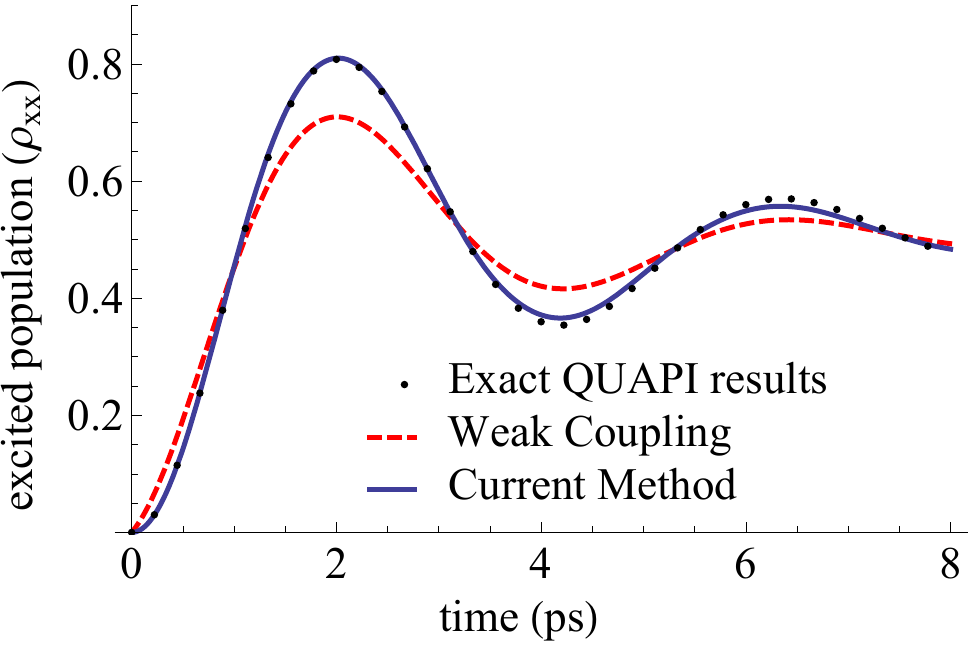}
\caption{
\label{Spin-Boson Figure1}
A comparison between the dynamics given by Eqn.~\ref{ansatz1} (solid), standard Born-Markov weak-coupling Master equation approach (dashed) given in Appendix \ref{appendix weakCoupling}, and exact QUAPI simulation of the model (dotted). For details of the calculations, see main text.
The parameters for this figure are taken from Ref.~\onlinecite{McCutcheon2011}:
$\Delta = \pi/2 \text{ ps}^{-1}$ , 
$\gamma(\omega) = 0$,
$\epsilon = 0$ ,
$T = 50 K$,
$J(\omega) = \alpha \omega^3 e^{-\omega^2/\omega_c^2}$,
$\alpha = 0.00675\text{ ps}^{-2}$, 
$\omega_c = 2.2\text{ ps}^{-1}$
.}
\end{figure}

A comparison between the standard Markovian Master equation, the current method and exact numerical simulation for the case of a super-Ohmic environment is shown in Fig.~\ref{Spin-Boson Figure1}. The QUAPI technique~\cite{Makri1992,Makri1995,Makri1995a} is used as an exact numerical benchmark curve: Our calculation uses nine kernel time steps, covering a total kernel memory time of 2~ps and is fully converged. The standard Born-Markov weak-coupling approach is given in Appendix \ref{appendix weakCoupling}. Clearly, our method's non--Markovian nature and lack of Born approximation results in an impressive improvement over the standard Born-Markov weak coupling ME approach. For this particular comparison, since there is no wider universe involved, $\gamma(\omega)=0$, the current method is equivalent to the second-order TCL approach, which also does not employ any approximations beyond a perturbation in the system-environment coupling. However, a key strength of the current formulation is that it is trivial to include a wider universe, which simply enters in the form of an exponential cut--off to the response function.

We note that this method allows us to easily study the case where the spectral density has several discrete sharp peaks as well as a smooth background, which is believed to be the case in many (if not all) systems studied in quantum biology~\cite{Kreisbeck2012,Nalbach2011b}. In this case the response function vanishes very slowly, which makes an exact numerical treatment extremely demanding, as a long history of the system needs to be tracked. In some papers, such as Ref.~\onlinecite{Kreisbeck2012} this issue is resolved by approximating a delta-function peak in the spectral density as a Lorentzian with a finite width. 
We note that if one allows this single peak to be damped, then in light of Eqn.~(\ref{effTemp eq}), this mode drives the system to an effective temperature different from the initial temperature of the environment $T$. Hence replacing discrete modes with Lorentzian distributions added to a continuous spectral density may in some parameters regimes become a questionable approximation.
By contrast, the additive property of modes to the influence functional $\Theta(t)$ here allows us to combine a discrete set of modes with a smooth background by taking
\begin{gather}
\Theta(t) = \Theta_\text{smooth}(t)+\Theta_\text{discrete}(t)~.
\end{gather}

As an example for this, let us study the spin boson model with a smooth background of oscillators plus a more strongly coupled discrete peak of frequency $\omega_s$ in the environment. We single out this peak and label it henceforth with a subscript $s$, writing the system-environment Hamiltonian as
\begin{align}\label{multi single hamiltonian}
\Hamiltonian_{SE} =& \half \epsilon \sigma_z + \half \Delta \sigma_x + \sum_k \omega_k \ad_k a_k + \omega_s \ad_s a_s \notag
\\&+ \sigma_z\left(\sum_k g_k (a_k+\ad_k)+g_s(\ad_s+a_s)\right).
\end{align}
\begin{figure}
\centering
\includegraphics[scale=0.8]{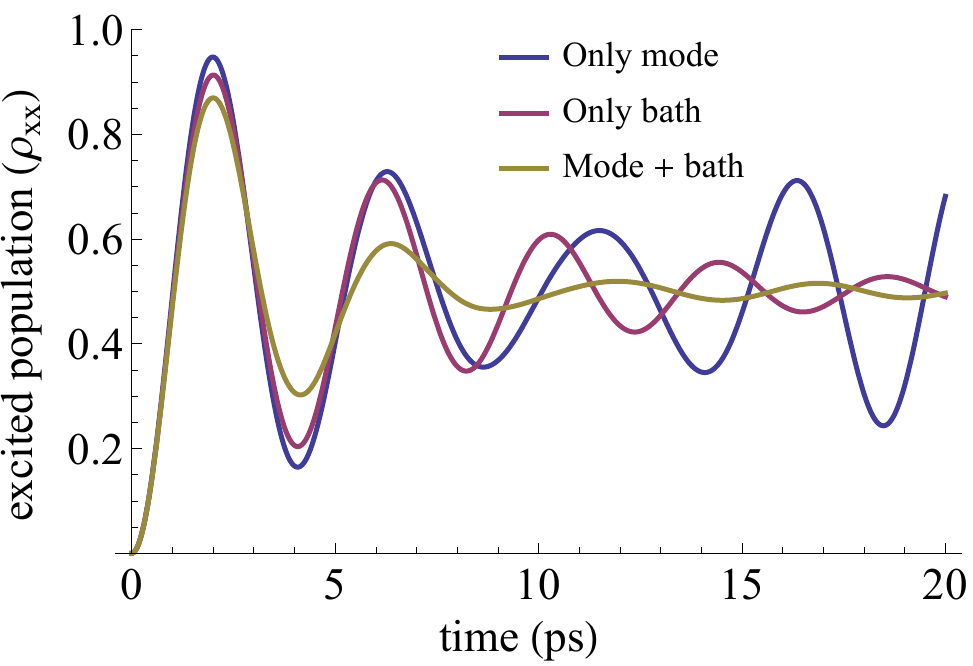}
\caption{
\label{multi single both Figure}
A comparison of quantum dynamics in a two level system that is coupled individually to a single mode, or to a continuous bath, or to a combination of the two.
The parameters used here are the same as the ones of Fig.~\ref{Spin-Boson Figure1}, but with a smaller coupling $\alpha = 0.0027~\text{ ps}^{-2}$, and with an added detuned single peak according to Hamiltonian (\ref{multi single hamiltonian}) with $g_s=0.1~\text{ps}^{-1}$, $\omega_s = 1.02 \Delta$. The mode is damped with rate $\gamma_s = 0.05~\text{ps}^{-1}$. 
}
\end{figure}
In Fig.~\ref{multi single both Figure} we start with the system in its ground state and plot the excited state population $\rho_{xx}$ as a function of time, for the cases where the system is only coupled to a smooth environment ($g_s\rightarrow 0$), only coupled to a single mode ($\{g_k\} \rightarrow 0$), and for the combined case.

\begin{figure}
\centering
\includegraphics[scale=0.9]{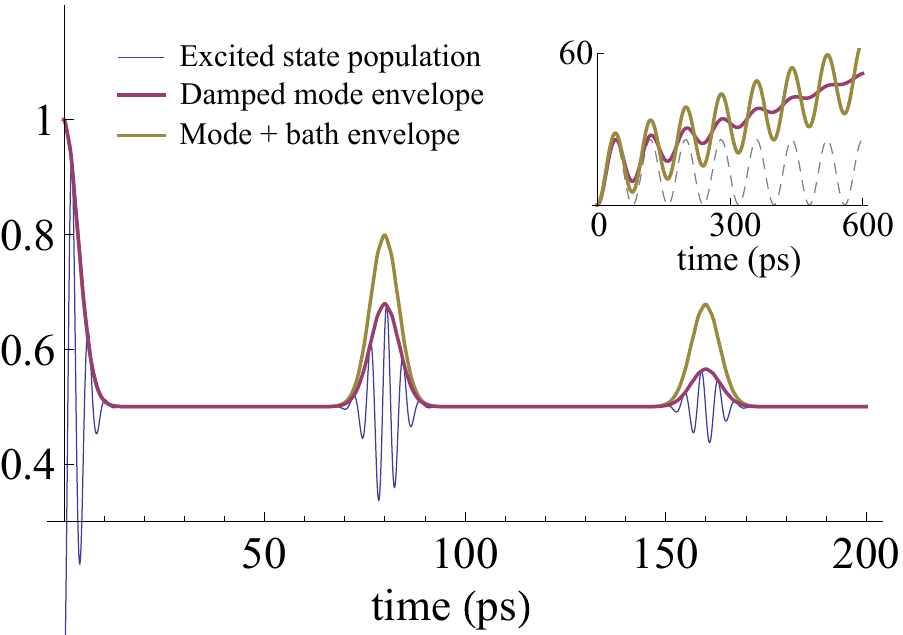}
\caption{
\label{figure Revivals}
Long time population of a TLS (blue), illustrating the revivals which occur when a discrete system is coupled to a single (damped) oscillator mode. The corresponding relaxation envelope (purple) and that of an undamped mode but where the system is coupled to a bath (yellow) are also shown. Here, we have chosen a bath coupling strength to obtain the same average relaxation rate for both cases (c.f.~inset), even though this does not become apparent during the first two revivals. The parameters in this figure are chosen to show revivals, so that the mode is almost resonant with the TLS and the damping is weak,
$\Delta = \pi/2~\text{ps}^{-1}$, 
$\gamma = 0.001~\text{ps}^{-1}$,
$\epsilon = 0$,
$\omega_s = 1.05 \Delta$,
$k_b T = 6.546~\text{ps}^{-1}$, and
$g_s = 0.1~\text{ps}^{-1}$.
The inset shows the relaxation exponent $-\theta_\text{relax}(t)$ with the same parameters as main figure but increased  $\gamma = 0.01~\text{ps}^{-1}$. Here it becomes apparent that the average gradient, i.e. average relaxation rate, is matched. The dashed curve of the inset is for reference, indicating the frequency of revivals by setting $\gamma=0$. 
}
\end{figure}

Due to the non-Markovian nature of this method, we are able to capture the revival effect~\cite{Approximations1984} for the Rabi model. These revivals can be damped via a combination of two mechanisms: Either the mode itself is coupled to a wider environment damping it, or there might be an additional continuous bath directly damping the system. In Fig.~\ref{figure Revivals} we plot the first case, where the environment consists of a single damped mode. The damping of the mode induces relaxation rate given by $\Gamma_1 = $ Eqn.~(\ref{relaxation constant eq}). We also plot the decay envelope $= \half+\half \exp \Theta_\text{relax}(t)$ for this case, as well as the decay envelope produced by coupling of the system to a continuous bath and no damping on the mode, choosing parameters such that the relaxation rate induced by the bath Eqn.~(\ref{dephase constant eq multimode}) is equal to $\Gamma_1$. This second decay envelope is then given by the expression $\half+\half \exp [\Theta^\text{single mode}_\text{relax}(t)+\Theta^\text{smooth}_\text{relax}(t)]$.
We note that the second case yields an exponential envelope to the dynamics for times $t \gg t_\text{relax}^R$, while for a single damped mode with damping rate $\gamma$, the envelope only becomes exponential for times $t \gg 1/\gamma$, which could be much longer. We note that the Lamb-shift given by Eqn.~(\ref{eq theta for spin-boson 2}) also differs between the two cases, albeit in the plotted parameter regime this difference is very subtle and not shown.

\section{Discussion and Conclusion}\label{summary}

We have introduced a novel method for studying a ubiquitous open quantum systems problem. Our approach differentiates between the immediate environment of the system of interest and a wider universe which effectively serves as a heat bath for this environment; this hierarchy of environments corresponds to many practical situations and is -- remarkably -- accomplished by a simple redefinition of the response kernel. The expressions resulting from our method are easy to evaluate numerically, and scale favourably with increasing system size. Moreover, the method still leads to soluble equations when the system of interest possesses a general time dependent Hamiltonian.
 
Whilst our method is limited to the weak coupling regime, it performs favourably when compared with traditional Born-Markov weak coupling master equations. Its approximate analytical expressions scale well with increasing system size and permit valuable physical insight, in contrast to some numerically exact approaches. Like many recent developments in the field of open systems, see~e.g.~Refs.~\onlinecite{Pollock2013, ilessmith2014, Higgins2013, Ritschel2011b} (and with the notable exception of Ref.~\onlinecite{Woods2014}), we do not presently have stringent criteria demarcating its precise regime of validity, which must thus be established by comparison with exact numerics. As a general guideline, however, our technique can be expected to perform well whenever other weak coupling approaches such as the time-convolutionless or the Nakajima-Zwanzig projection operator expansions\cite{breuer2007theory} are valid for the system-to-immediate-environment coupling. As an additional criterion, our treatment of the  wider universe (if present) assumes that $\gamma_k \ll \omega_k$, i.e.~each mode is weakly coupled to its heat bath.

We have benchmarked our technique against the well-studied spin boson model and the Rabi model, finding it leads to expressions that are indeed highly accurate when compared with numerically converged solutions. This remains true even for coupling strengths where a conventional standard second order Born Markov master equation begins to performs poorly, and exactly recovers the time-convolutionless solution when no wider universe is present. For cases when the system-environment coupling is not sufficiently weak for the second order expansion of the interaction, we provide an explicit recipe to calculate higher orders in the perturbation series. Perhaps a unique advantage of this approach is that these two models, i.e. the Rabi and the spin-boson models can easily be combined even for long-time dynamics. This makes our method eminently suitable for studying the exciton energy transfer in photosynthetic or artificial molecular systems, since the coupling of the excitonic degree of freedom to both the vibrational quasi-continuum of the wider protein scaffolding as well as to specific localised vibronic modes is believed to be of crucial functional importance.

\begin{acknowledgements}
We thank Elinor Irish, Kieran Higgins and Elliott Levi for stimulating discussions. 
This work was supported by the Leverhulme Trust, EPSRC under platform grant EP/J015067/1, and the National Research Foundation and Ministry of Education, Singapore. BWL thanks the Royal Society for a University Research Fellowship. EMG acknowledges support from the RSE/Scottich government.
\end{acknowledgements}

\appendix

\section{Multiple Modes}\label{appendix multimode}
We start from Hamiltonian (\ref{Hamiltonian1}) and Eqns.~(\ref{multi mode Hamiltonian 1}-\ref{multi mode Hamiltonian 2}), and look at the case where all of the modes are coupled in the same manner (same $V$ operator) but with different strengths $g_k$. For multiple modes the density matrix is represented by Eqn.~(\ref{general den matrix}), and
the operator correspondence between $\rho$ and $\vec{P}$ is:
\begin{align}
\pd{}{t}\rho =& -i[H,\rho] + D(\rho) \leftrightarrow \\
\pd{}{t}\vec{P} =& -i(\Hamiltonian_S^\times+L) \vec{P} + \sum_k g_k A_g(k) \vec{P} ~, \label{P correspondence multi}
\end{align}
where now
\begin{align}
L = \sum_k\Big[&\left(-\omega_k+\frac{i}{2}\gamma_k\right)\dalpha_k\alpha_k + \left(\omega_k+\frac{i}{2}\gamma_k\right)\dalphas_k\alphas_k 
\notag \\
&+ i\gamma_k N_k  \pd{^2}{\alpha_k\partial \alphas_k}\Big] ~,
\end{align}
$N_k = (e^{\beta \omega_k}-1)^{-1}$ and $\gamma_k = \gamma(\omega_k)$ is the damping rate of mode $k$. The matrices $A_g(k)$ are given by
\begin{align}
\left[A_g(k)\right]_{ij} =& -i\left(\pd{}{\alphas_k}-\pd{}{\alpha_k}\right)[V_l(t) d_{lij}+V_{n^2}(t)\delta_{ij}] \notag
\\-\Big(2\alpha_k&+2\alphas_k-\pd{}{\alpha_k}-\pd{}{\alphas_k}\Big)V_l(t) f_{lij} ~, 
\label{agk1}
\\
\left[A_g(k)\right]_{i, n^2} =& -i\left(\pd{}{\alphas_k}-\pd{}{\alpha_k}\right)V_i(t) ~, \\
\left[A_g(k)\right]_{n^2, i} =& -i\left(\pd{}{\alphas_k}-\pd{}{\alpha_k}\right)\frac{2}{n}V_i(t) ~, \\
\left[A_g(k)\right]_{n^2, n^2} =& -i\left(\pd{}{\alphas_k}-\pd{}{\alpha_k}\right)V_{n^2}(t) ~.
\label{agk4}
\end{align}
Assuming all of the couplings $g_k$ are sufficiently small, at the order of $\sum_k g_k \sim g$, we can rewrite Eqn.~(\ref{P correspondence multi}) to become
\begin{gather}\label{fullequation nsites multmode}
\pd{}{t}\vec{P} = -i(\Hamiltonian_S^\times+L) \vec{P} + g \left(\sum_k \tilde{g_k} A_g(k)\right) \vec{P}
\end{gather}
with $g_k = g \tilde{g_k}$. Now consider the perturbative expansion
\begin{gather}\label{pertur nsites multi}
P = P^0 + g P^1 + g^2 P^2 + \cdots ~,
\end{gather}
so that Eqn.~(\ref{fullequation nsites multmode}) translates to:
\begin{align}
\pd{}{t} P^0 &= -i(\Hamiltonian_S^\times+L) P^0 \label{P0 eq nsites multi} ~, \\
\pd{}{t} P^1 &= -i(\Hamiltonian_S^\times+L) P^1 + \sum_k \tilde{g_k} A_g(k) P^0 \label{P1 eq nsites multi} ~, \\
\pd{}{t} P^2 &= -i(\Hamiltonian_S^\times+L) P^2 + \sum_k \tilde{g_k} A_g(k) P^1 ~, \\
\cdots\notag
\\
\pd{}{t} P^n &= -i(\Hamiltonian_S^\times+L) P^n + \sum_k \tilde{g_k} A_g(k) P^{n-1} ~. \label{Pn eq nsites multi}
\end{align}
The solution for the uncoupled system $P^0$ is then equivalent to the single mode case, and is given by (assuming a factorized initial state):
\begin{gather}
P^0(t) = U(t)\rho^s(0)
\prod_k \frac{1}{\pi N_k}e^{-|\alpha_k|^2/N_k} ~.\label{multi thermal state}
\end{gather}
We assume that $\alpha_k^l P^n(\alpha)\underset{\alpha\rightarrow\infty}\longrightarrow 0$ for all $k,n,l$ for the same reasons given in the main text.  Performing the integration $\int_{\alpha}$ on Eqns.~(\ref{P1 eq nsites multi}-\ref{Pn eq nsites multi}) yields
\begin{gather}
\pd{}{t} \int_\alpha P^1 = -i\Hamiltonian_S^\times \int_\alpha P^1 -i \sum_k \tilde{g_k}V^\times\underbrace{\int_\alpha (\alpha_k+\alphas_k)P^0}_{\rightarrow 0} \label{P1 inteq multi} ~, \\
\pd{}{t} \int_\alpha P^2 = -i\Hamiltonian_S^\times \int_\alpha P^2 -i \sum_k \tilde{g_k} V^\times\int_\alpha (\alpha_k+\alphas_k)P^1 \label{P2 inteq nsites multi} ~, \\
\cdots \nonumber ~, \\
\pd{}{t} \int_\alpha P^n = -i\Hamiltonian_S^\times \int_\alpha P^n -i \sum_k \tilde{g_k} V^\times\int_\alpha (\alpha_k+\alphas_k)P^{n-1} ~, \label{Pn inteq nsites multi} 
\end{gather}
where just as before, $V^\times$ is the equivalent of $[V,\Box]$ and is given by Eqn.~(\ref{A_1 def}), and the initial condition is 
$\int_\alpha P^{n>0}(t=0) = 0$ , i.e. at time $t=0$ the system and the environment were factorized. The first contribution in the expansion comes from $\int_\alpha P^2\neq 0$, which is $2^{nd}$ order in the coupling constant $g$. 
In order to solve Eqn.~(\ref{P2 inteq nsites multi}) we first need to evaluate the expression $\int_{\alpha} (\alpha_k+\alphas_k) P^1$ for each $k$, which is accomplished by multiplying Eqn.~(\ref{P1 eq nsites multi}) by $\alpha_{k'}$ or $\alphas_{k'}$ from the left, and then performing the $\int_\alpha$ integral. As a consequence, all of the terms in the sum with index $k\neq k'$ vanish, and we are left with
\begin{gather}
\pd{}{t}\int_\alpha \alpha_k P^1 = -i\int_\alpha \alpha_k (\Hamiltonian_S^\times+L) P^1 + \tilde{g_k} \int_\alpha \alpha_k A_g(t,k) P^0
\end{gather}
and a corresponding equation for $\alphas_k$. Crucially, there is no sum over $k$ here, which means each $k$ gives rise to exactly two equations of the type of Eqns.~($\ref{originaltwoeq gen}$, $\ref{originaltwoeq gen2}$), which we have already solved. The first non-vanishing term is hence given by
\begin{gather}\label{p2multi}
\int_\alpha P^2 = -U(t) \int_0^t dt' \int_0^{t'} dt'' \tilde{V}^\times(t')\times
\\
\sum_k \tilde{g}_k^2 e^{-\half \gamma_k(t'-t'')}\Big[
(2N_k+1)\cos[\omega_k(t'-t'')] \tilde{V}^\times(t'') \notag\\
-i\sin[\omega_k(t'-t'')]\tilde{V}^\circ(t'')\Big] \rho^s(0)\notag ~,
\end{gather}
which is just Eqn.~(\ref{intalphaP2}) with an added sum over all modes, and where $\tilde{V}^{\times,\circ}$ are given by Eqn.~(\ref{a1a2 def}). From here we continue to Eqn.~(\ref{eq Theta multimode}).

\section{Higher Orders Calculation}\label{appendix higherOrders}
In this Appendix, we show how to calculate higher orders of the influence functional $\Theta(t)$ defined in Eqn.~\ref{ansatz1}, where the main text only gives the 2$^\text{nd}$ order expression.
We also show that, in analogy to the known result of the non-hierarchichal case~\citep{breuer2007theory}, all of the odd orders vanish when the initial state factorises $\rho(0) = \rho_s(0)\otimes\rho_E^{th}$.

We start by giving a formal expression of the quantity
\begin{gather}
U(t)\chi_n\big(\{a_i\},\{b_i\};t\big)\rho^s(0) \equiv \int_\alpha \prod_i (\alpha_i)^{a_i} (\alpha_i^{*})^{b_i} P^n~,\label{highOrder gen}
\end{gather}
where $\{a_i\}$ and $\{b_i\}$ are non-negative integers. Begin by multiply Eqn.~(\ref{Pn eq nsites multi}) by $\prod_i (\alpha_i)^{a_i} (\alpha_i^{*})^{b_i}$ and integrate over $\alpha$ to obtain
\begin{widetext}
\begin{align}
\Big(\pd{}{t} +i \Hamiltonian_S^\times + &\sum_k [i\omega_k(a_k-b_k)+\half\gamma_k (a_k+b_k)] \Big)\times 
\int_\alpha \prod_i (\alpha_i)^{a_i} (\alpha_i^{*})^{b_i} P^n = 
\notag\\
& = \sum_k \gamma_k N_k a_k b_k \int_\alpha \prod_i (\alpha_i)^{a_i} (\alpha_i^{*})^{b_i} \frac{1}{\alpha_k \alphas_k} P^n + \sum_k \tilde g_k \int_\alpha \prod_i (\alpha_i)^{a_i} (\alpha_i^{*})^{b_i}A_g(k) P^{n-1}\label{highOrderODE}~,
\end{align}
which gives
\begin{gather}
\chi_n\big(\{a_i\},\{b_i\};t\big) = \int_0^t d\tau e^{-\sum_k [i\omega_k(a_k-b_k)+\half\gamma_k (a_k+b_k)](t-\tau)}S_n(\{a_i\},\{b_i\};\tau)
\label{highOrder chi}
\end{gather}
where $U(\tau) S_n(\{a_i\},\{b_i\};\tau)\rho^s(0)$ is the RHS of Eqn.~(\ref{highOrderODE}). 
Using the definition of $A_g$ [Eqns. (\ref{agk1}-\ref{agk4})]
we get the following expressions:
\begin{align}
S_n(\{a_i\},\{b_i\};t) =& \sum_k \gamma_k N_k a_k b_k \chi_n(a_k-1,b_k-1;t) 
\label{highOrder sn}
\\
& - i\tilde{V}^\times(t)\sum_k \tilde{g}_k\left[\chi_{n-1}(a_k+1;t)+\chi_{n-1}(b_k+1;t)+\frac{a_k}{2}\chi_{n-1}(a_k-1;t)+\frac{b_k}{2}\chi_{n-1}(b_k-1;t)\right]
\notag\\
& -\frac{i}{2} \tilde{V}^\circ(t)\sum_k \tilde{g}_k\left[a_k\chi_{n-1}(a_k-1;t)-b_k\chi_{n-1}(b_k-1;t)\right].
\notag
\end{align}

\end{widetext}
In the above expression we used $\tilde{V}(t)$ which is defined by Eqn.~(\ref{a1a2 def}), and the sloppy notation
$
\chi_n(a_k-1;t) = \chi_n(a_1,\cdots,a_k-1,\cdots,b1,b2,\cdots ; t)
$. Complemented by the initial condition
\begin{gather}
\label{highOrder initial}
\chi_0\big(\{a_i\},\{b_i\}\big) = 
\begin{cases}
\prod_k (a_k !) (N_k)^{a_k} & \forall i , a_i=b_i
\\
0 & \text{else}
\end{cases}
\end{gather}
we can in principle get the expression for Eqn.~(\ref{highOrder gen}).
From examination of Eqns.~(\ref{highOrder chi},\ref{highOrder sn},\ref{highOrder initial}) it is evident that if $n+\sum_i(a_i+b_i)$ is odd, then
\begin{gather}
\int_\alpha \prod_i (\alpha_i)^{a_i} (\alpha_i^{*})^{b_i} P^n = 0.
\end{gather}
This means that in the series $\Theta(t) = \sum_i g^i \Theta_i(t)$, all odd powers of $g$ vanish. Finally, we can express the influence functional $\Theta(t)$ as
\begin{gather}
\Theta(t) = g^2 \Theta_2 + g^4 \Theta_4 + g^6 \Theta_6 + \cdots
\end{gather}
with
\begin{gather}
\Theta_2 = \chi_2(0;t) ~,
\\
\Theta_4 = \chi_4(0;t) - \frac{\Theta_2^2}{2!}~,
\\
\Theta_6 = \chi_6(0;t) - \frac{\Theta_2 \Theta_4 + \Theta_4 \Theta_2}{2!} - \frac{\Theta_2^3}{3!}~,
\\
\Theta_8 = \chi_8(0;t) - \frac{\Theta_4 \Theta_4 + \Theta_6 \Theta_2 + \Theta_2 \Theta_6}{2!} \\ \notag - \frac{\Theta_2\Theta_4\Theta_4+\Theta_4\Theta_2\Theta_4+\Theta_4\Theta_4\Theta_2}{3!}
- \frac{\Theta_2^4}{4!}~,
\end{gather}
etc. Here we used $\chi(0;t) = \chi(\{a_i=0\},\{b_i=0\};t)$.
\section{Standard Born-Markov Weak-Coupling Master Equation}\label{appendix weakCoupling}
In this Appendix, we follow the recipe given in chapter 3 of Ref.~\onlinecite{breuer2007theory} in order to derive the standard Born-Markov weak-coupling master equation that is one of our benchmarks throughout the paper. We start from the Rabi Hamiltonian given by Eqn.~(\ref{Rabi Hamiltonian}), ignoring $\Hamiltonian_{EU} = 0$ for now.
With the suitable change of basis we can write this Hamiltonian as
\begin{gather}
\tilde{\Hamiltonian}_R = \frac{\Omega}{2}\tilde\sigma_x+\omega \ad a + \frac{g}{\Omega}[\epsilon \tilde\sigma_x+\Delta(\tilde\sigma_++\tilde\sigma_-)](\ad+a)~,
\end{gather}
where the tilde denotes the new basis, $\tilde\sigma_\pm$ are the lowering and raising operators, and $\Omega = \sqrt{\epsilon^2+\Delta^2}$ is the Rabi frequency. Adopting the notation from Ref.~\onlinecite{breuer2007theory}, we have
\begin{gather}
A(\pm\Omega) = g\frac{\Delta}{\Omega}\tilde\sigma_\pm ~,\\
A(0) = g\frac{\epsilon}{\Omega}\tilde\sigma_x ~,\\
S(\alpha) = \frac{N(\omega)}{\alpha+\omega}+\frac{N(\omega)+1}{\alpha-\omega} ~,\\
\gamma(\alpha) = \frac{\pi}{2}\delta(\alpha+\omega) N(\omega) + \frac{\pi}{2}\delta(\alpha-\omega) [N(\omega)+1] ~.
\end{gather}
This defines the Lamb-Shift Hamiltonian as
\begin{align}
\tilde\Hamiltonian_\text{LS} &= \sum_{\alpha=0,\pm\Omega} S(\alpha)A(\alpha)A^\dagger(\alpha) 
\\ &= g^2\frac{\Delta^2}{\Omega^2}\frac{\Omega}{\Omega^2-\omega^2}\coth(\frac{\beta \omega}{2})\tilde\sigma_x ~,\label{eq appendix lambshift}
\end{align}
up to a constant that does not affect the dynamics. The dissipator is given by
\begin{align}
D(\rho_s) =& \\
\sum_{\alpha=0,\pm\Omega} &\gamma(\alpha)\left(A(\alpha) \rho_s A^\dagger(\alpha)-\half\{A^\dagger(\alpha)A(\alpha),\rho_s\} \right)
\notag\\
=&~g^2 \frac{\Delta^2}{\Omega^2}\frac{\pi}{2}\delta(\Omega-\omega) \times \\
&~~\Big[(N(\omega)+1)(\tilde\sigma_+ \rho_s \tilde \sigma_- - \half\{\tilde\sigma_-\tilde\sigma_+,\rho_s\})
\notag\\
&~~~+N(\omega))(\tilde\sigma_- \rho_s \tilde \sigma_+ - \half\{\tilde\sigma_+\tilde\sigma_-,\rho_s\})\Big]
\notag\\
&~~+g^2 \frac{\epsilon^2}{\Omega^2}\frac{\pi}{2}\delta(\omega)\coth\left(\frac{\beta \omega}{2}\right)(\tilde\sigma_x\rho_s\tilde\sigma_x-\rho_s)\notag ~,
\end{align}
and the dynamics of the system is then governed by
\begin{gather}
\pd{}{t}\rho_s = -i[\tilde\Hamiltonian_R + \tilde\Hamiltonian_\text{LS},\rho_s] + D(\rho_s)~.
\end{gather}\label{eq appendix WCME}
From the above expression we can extract the relaxation and dephasing rates, obtaining
\begin{gather}\label{eq appendix relaxation}
\Gamma_\text{relax} = 2\pi g^2 \coth\left(\frac{\beta \Omega}{2}\right) \frac{\Delta^2}{\Omega^2} \delta(\Omega-\omega) ~,
\\
\label{eq appendix dephasing}
\Gamma_\text{dephase} = \frac{1}{2}\Gamma_\text{relax} + 4\pi g^2 \coth\left(\frac{\beta \omega}{2}\right)\frac{\epsilon^2}{\Omega^2}\delta(\omega)~.
\end{gather}
At this point we can easily calculate the relaxation and dephasing rates, as well as the Lamb-shift Hamiltonian for the spin-boson Hamiltonian from Eqn.~(\ref{eq Spin-Boson Hamiltonian}), simply but summing over the contributions from each mode of the bath. In terms of the spectral density Eqn.~(\ref{eq Spectral density}), the rates are then given by
\begin{gather}
\tilde\Hamiltonian_\text{LS} = \tilde\sigma_x\frac{\Delta^2}{\Omega^2}\int_0^\infty d\omega J(\omega) \coth\left(\frac{\beta \omega}{2}\right)\frac{\Omega}{\Omega^2-\omega^2}~,
\\
\Gamma_\text{relax} = 2\pi \frac{\Delta^2}{\Omega^2} J(\Omega)\coth\left(\frac{\beta \Omega}{2}\right) ~,
\\
\Gamma_\text{dephase} = \half\Gamma_\text{relax} +4\pi \frac{\epsilon^2}{\Omega^2} k_b T \lim_{\omega\rightarrow 0}\frac{J(\omega)}{\omega}~.
\end{gather}
\vfill
\bibliography{biblio}

\end{document}